\documentclass[structabstract]{aa}  
\usepackage{graphicx}
\begin{document}
   \title{On the relationship of shock waves to flares and coronal mass 
ejections}

   \author{A. Nindos
          \inst{1}
          C.E. Alissandrakis
          \inst{1}
          A. Hillaris
          \inst{2}         
          \and
          P. Preka-Papadema\inst{2}
          }

   \institute{Section of Astrogeophysics, Physics Department, University of
              Ioannina, Ioannina GR-45110, Greece\\
              \email{anindos@cc.uoi.gr}\\
             \and
             Section of Astronomy, Astrophysics and Mechanics, Physics 
Department, University of Athens, Athens GR-15783, Greece\\
             }

   \date{Received ...; accepted ...}

 
  \abstract
   {Metric type II bursts are the most direct diagnostic of shock waves
in the solar corona.}
   {There are two main competing views about the origin of coronal shocks:
that they originate in either blast waves ignited by the pressure pulse of a 
flare or piston-driven shocks due to coronal mass ejections (CMEs). We 
studied three well-observed type II bursts in an attempt to place tighter
constraints on their origins.}
{The type II bursts were observed by the ARTEMIS radio spectrograph and
imaged by the Nan\c{c}ay Radioheliograph (NRH) at least at two frequencies.
To take advantage of projection effects, we selected events that 
occurred away from disk center.}
{In all events, both flares and CMEs were observed. In the first event, the 
speed of the shock was about 4200 km/s, while the speed of the CME was 
about 850 km/s. This discrepancy ruled out the CME as the
primary shock driver. The CME may have played a role in the ignition
of another shock that occurred just after the high speed one. A CME
driver was excluded from the second event as well because the CMEs
that appeared in the coronagraph data were not synchronized with the
type II burst. In the third event, the kinematics of the CME which was
determined by combining EUV and white light  data was broadly
consistent with the kinematics of the type II burst, and, therefore, 
the shock was probably CME-driven.}
   {Our study demonstrates the diversity of conditions that may lead
to the generation of coronal shocks.}

\authorrunning{Nindos et al.}
\titlerunning{Shock waves, flares, and CMEs}
   \maketitle
%

\section{Introduction}

The dynamic spectra of type II radio bursts appear as slowly drifting
bands,  from high to low frequencies. Most type II bursts appear
typically below 150 MHz (Maxwell and Thompson \cite{Maxwell62}; Mann
et al. \cite{Mann96}),  but occasionally the starting frequency may be
as high as 500 MHz (Pohjolainen et al. \cite{Pohjolainen08}). The
frequency drift rates of type II bursts are on the order of 0.3 MHz/s
and are found to increase with increasing starting frequency in the
meter wavelength range (e.g. Mann et al. \cite{Mann95};
\cite{Mann96}). Under the assumption that the exciter propagates along
the gradient of a typical coronal density model, the drift rate yields
speeds of 500-1000 km/s, whereas the Alfv\'{e}n speed in the low
corona is thought to be typically several hundred kilometers per
second; this led to the idea that metric type II bursts are generated
by  coronal MHD shocks travelling at several times the local
Alfv\'{e}n speed. The radio emission is  generated at the local plasma
frequency and/or its second harmonic; it is produced by either
energetic electrons accelerated at the shock front or plasma
turbulence excited by the shock. Detailed discussions of the
properties of type II bursts are given by Nelson and  Melrose
(\cite{Nelson85}), Vr\v{s}nak and Cliver (\cite{Vrsnak08}), Nindos et
al. (\cite{Nindos08}), and Pick and Vilmer (\cite{Pick08}).

\begin{table*}
\caption{Parameters of flares and CMEs}
\centering
\begin{tabular}{lcccc}
\hline
Event                          & Location  & GOES  & Soft X-rays  & CME speed \\
                               &           & class & Start Peak   &  (km/s)  \\
\hline
March 2, 2000                  &  S19W60   & M6.5  & 13:35 13:43  &  850 \\
March 7, 2000                  &  S13W60   & C6.3  & 14:21 14:30  &  -    \\ 
May 2, 2000                    &  N22W68   & M2.8  & 14:42 14:51  &   1278  \\
\hline
\end{tabular}
\end{table*}

\begin{table*}
\caption{Parameters of type II bursts}
\centering
\begin{tabular}{lcccccc}
\hline
Event                          & Frequency range & Start-end       & Drift rate                             &Type II onset     & NRH beam \\
                               &     (MHz)       &   time (UT)     & ($\frac{1}{f}\frac{df}{dt}$, s$^{-1}$) &    (UT)          & ($\arcsec$, 236 MHz) \\
\hline
March 2, 2000\tablefootmark{a} &  154-274   & 13:41:01-13:41:27 & -0.020                                 &  13:40:03        & $155 \times 254$ \\
March 2, 2000\tablefootmark{b} &  7.8-70X   & 13:41:40-13:57:00 &  -0.001\tablefootmark{c}               &13:20:25-13:25:27 &  - \\
March 7, 2000                  &  119-249   & 14:25:50-14:26:53 & -0.008- -0.010                             &14:24:12-14:24:50 & $286 \times 148$    \\
May 2, 2000                    &  127-340   & 14:48:04-14:52:53 &  -0.005                                   &14:45:10          &    $208 \times 135$ \\
\hline
\end{tabular}
\tablefoot{
\tablefoottext{a}{High-frequency type II burst}\\
\tablefoottext{b}{Low-frequency type II burst}\\
\tablefoottext{c}{F2a lane}
}
\end{table*}

While there is a concensus that coronal mass ejections (CMEs) are the
drivers of interplanetary shocks (e.g. Gopalswamy
\cite{Gopalswamy06}), the origin of coronal shock waves is not
completely understood.  Two possible physical interpretations have
been proposed: a blast wave ignited by the pressure pulse of a flare,
or, alternatively, a piston-driven shock due to a CME.  Small-scale
ejecta are also considered as a possible material driver of the shock
(e.g. Gopalswamy et al. \cite{Gopalswamy98}; Klein et
al. \cite{Klein99}; Klassen et al. \cite{Klassen03}; Dauphin et
al. \cite{Dauphin06}). These structures act as temporary pistons,
generating an initially driven shock. After the expansion stops, the
disturbance travels as a blast shock.  Statistical studies reveal that
most, if not all, metric type II bursts occur during events where both
flares and CMEs are observed. The appearance of metric type II bursts
without CMEs is exceptionally rare. On the other hand, whenever a
metric type II burst without a flare is observed, the flare probably
occurred behind the limb.

\begin{figure*}
\centering
\includegraphics[width=15cm]{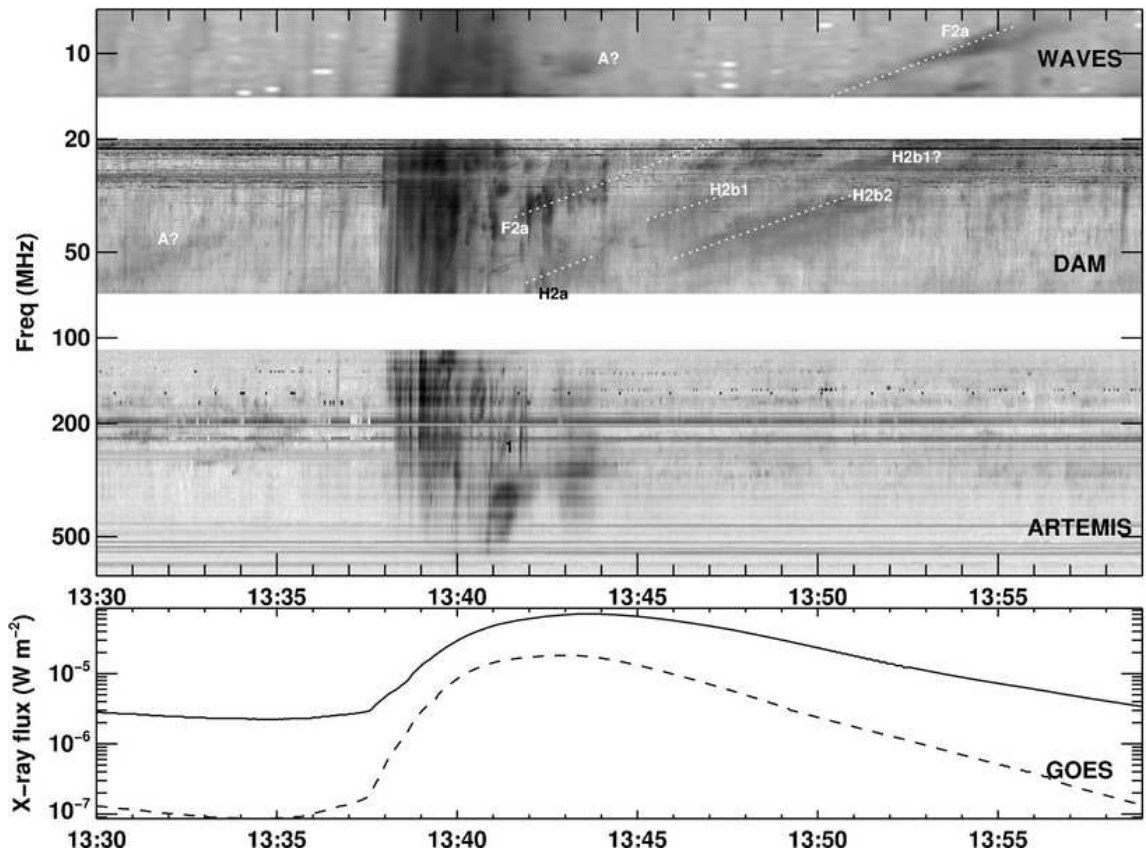}
\caption{{\it Upper panel:} Composite dynamic spectrum of the 2000
March  2 event observed between 687 to 7 MHz. From top to bottom,
the  Wind/WAVES (7-13.825 MHz), Nan\c{c}ay DAM (20-70 MHz),
and ARTEMIS data (110-687 MHz) are presented. 
{\it Lower panel:} Time profile of the GOES X-ray flux in the 1-8 \AA\ 
(solid line) and 0.5-4 \AA\ (dashed line)  energy channels.}
\end{figure*}

There have been many studies of the possible association of type II
bursts  with either flares or CMEs (e.g. Classen and Aurass
\cite{Classen02}; Shanmugaraju et al.  \cite{Shanmugaraju06}; Hudson
and Warmuth \cite{Hudson04}; Cho et al. \cite{Cho05}; \cite{Cho07};
Pohjolainen and Lehtinen \cite{Pohjolainen06}; Reiner et
al. \cite{Reiner07}; Vr\v{s}nak et al. \cite{Vrsnak06}; Dauphin et
al. \cite{Dauphin06}; Liu et al. \cite{Liu09}). Most type II bursts
that lack a clear CME association occur with flares within 30\degr\ of
central meridian where CMEs are difficult to observe (e.g. Classen and
Aurass \cite{Classen02}).   The traditional argument in favor of a
flare origin of coronal shocks is based  on the back extrapolation of
the type II lanes, which usually corresponds to the interval between
the onset and the peak energy release of the flare (Harvey
\cite{Harvey65}; Klassen et al. \cite{Klassen99}; Vr\v{s}nak
\cite{Vrsnak01a}). However, the flare impulsive phase is often closely
associated with the acceleration phase of the  flare-related CME
(e.g. Zhang et al. \cite{Zhang01}; Zhang et al. \cite{Zhang04}; Temmer
et al. \cite{Temmer08}).

Few comparisons between type II burst imaging and white-light CMEs
have been reported because the heights from which the metric emission
originates are usually occulted in space-borne coronagraphs. A
direct comparison is also difficult because electron acceleration at
shocks may be restricted to quasi-perpendicular regions such that some
difference between spatially restricted sources of type II emission
and the CME is expected. When simultaneous images are available, the
type II burst usually appears lower than the CME front (e.g. Gary et
al. \cite{Gary84}; Klein et al. \cite{Klein99}), even though in some
cases the shock source is located at or near the CME front (e.g. Maia
et al. \cite{Maia00}).

The strongest support of the CME-driven-shock scenario is probably
that provided by observations of broadening and intensity changes in
the UV emission lines in front of CMEs, which have been attributed to
shocks temporally associated with type II bursts (Raymond et
al. \cite{Raymond00}; Mancuso et al. \cite{Mancuso02}; Ciaravella et
al. \cite{Ciaravella05}). In addition, though less direct, support is
provided by the scaling of CME kinetic energy with the frequency range
of type II emission and the high association of EUV waves with
CMEs. Gopalswamy (\cite{Gopalswamy06}) and  Gopalswamy et
al. (\cite{Gopalswamy05}; \cite{Gopalswamy09})   suggested that CMEs
are the drivers of shocks producing metric type II emission. They
pointed out that an Alfv\'{e}n speed profile with a local maximum in
the corona could explain the discontinuity between type II bursts at
meter and longer wavelengths, even if they were generated by the same
CME. However, the analysis by Cane and Erickson (\cite{Cane05})
challenges the argument that coronal and interplanetary type II bursts
are related to the same driver.

\begin{figure}[h]
\centering
\includegraphics[width=9cm]{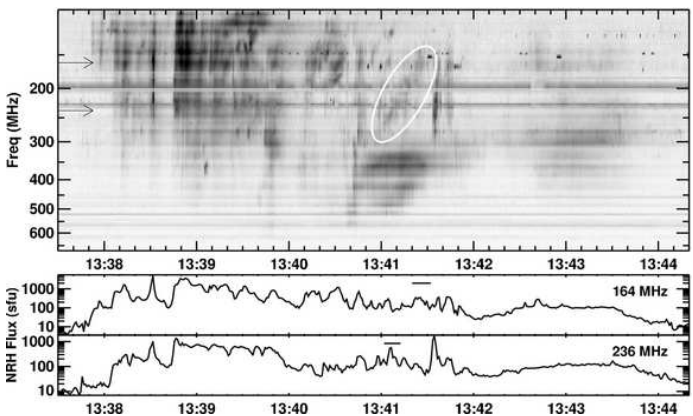}
\caption{{\it Upper panel:} Enlarged view of the dynamic spectrum of
the 2000 March 2 event  observed by ARTEMIS. The type II burst is surrounded
by the white ellipse. The arrows indicate the
type II frequencies (236 and 164 MHz) observed by the
NRH. {\it Lower panel:} Time profiles of the radio flux at 236 and 164
MHz derived from the NRH maps. The horizontal lines mark the passage
of the type II burst.}
\end{figure}

The most compelling support of a flare interpretation of type II
bursts was provided by both White et al. (\cite{White07}) and
Magdaleni\'{c} et al.  (\cite{Magdalenic08}; \cite{Magdalenic10}). 
White et al. (\cite{White07}) presented a type II burst that was 
associated with a long-duration CME-less X3.6-class flare. Magdaleni\'{c} 
et al. (\cite{Magdalenic08}; \cite{Magdalenic10}) considered a limb 
event and using radio and EUV images showed that the impulsive CME 
acceleration in their event lagged the type II burst that was synchronized 
with the associated flare.

Statistical studies (and in some cases studies of individual events)
often do  not yield unambiguous conclusions about the origin of type
II bursts. The aim of our article is to study the origin of three
well-observed type II bursts.  We selected events that occurred away
from the disk center (thus taking advantage of projection effects) and for
which, in addition to the spectral radio data,  positional information
about the shock radio sources were available from  radioheliograph
data. Our data are presented in section 2. The three events are
described in section 3. Our conclusions are summarized in section 4.
   

\section{Observations and data reduction}

We selected three events (see Table 1) that contained type II bursts
(see Table 2) with the following  criteria: (1) The events were
observed by the ARTEMIS radio spectrograph and  included in the
catalog of type II events presented by Caroubalos et
al. (\cite{Caroubalos04}). (2) The type II emission was imaged by the
Nan\c{c}ay Radioheliograph (NRH) at least at two frequencies. The
positional information helped us to remove (at least partially) the
ambiguity associated with the determination of the shock heights using
the type II drift rates and coronal density models. (3) The events
occurred at longitudes equal to or larger than 60\degr\ to take
advantage of projection effects.  Away from disk center, the
determination of the kinematics of both radio sources and CMEs becomes
more accurate, and CMEs are also easier to observe.

The ARTEMIS solar radio spectrograph is located at Thermopylae, Greece
(Caroubalos et al. \cite{Caroubalos01}). At the times of the events
analyzed in this article, the instrument covered the frequency range
from 110 to 687 MHz, using two receivers operating in parallel. In our
study, we used data from the global spectral analyzer (ASG), which is
a sweep frequency receiver that covered the full frequency band with a
time resolution of 10 samples s$^{-1}$.  The NRH (Kerdraon and Delouis
\cite{Kerdraon97}) is located 200 km south of Paris. At the times of
our observations, it consisted of 44 antennas, spread over two arms
(east-west and north-south) with respective lengths of 3200 m and 2440
m. It provided images of the full solar disk at five frequencies (164,
236, 327, 410, and 432 MHz) with sub-second time resolution. For the
observations reported here, the resolution of the images at 236 MHz is
given in Table 2. The extension of the type II burst emission to lower
frequencies than the ones covered by ARTEMIS was studied by using data
from the Nan\c{c}ay Decametric Array (DAM; 20-70 MHz) and the
WIND/WAVES RAD 2 receiver (1.075-13.825 MHz).

In addition to the radio data we used EUV observations from  the
Extreme Incidence Telescope (EIT) at 195 \AA\ and white-light
coronagraph data from Large Angle Spectroscopic COronagraph (LASCO) C2
and C3 coronagraphs. For the overall temporal evolution of the flares,
we used GOES X-ray total flux measurements.

\section{Results}

\subsection{The 2000 March 2 event}

The event of 2000 March 2 was related to a GOES M6.5-class flare that
occurred at 19\degr S 60\degr W (NOAA AR8882). A CME was also observed
in the LASCO images. The flare started at 13:35 UT and attained its
maximum at 13:43 UT (see Table 1). Before the M6.5-class flare, a
C5.5-class flare occurred between 13:06 and 13:26 UT in the same active
region. The C5.5-class flare peaked at 13:15 UT. The NRH 10-s
resolution data did not show any emission associated with the active
region prior to the event. This may be due to its insufficient dynamic
range because at the same time a strong compact  source close to disk
center was present.

Spectral radio observations are presented in the composite dynamic
spectrum of figure 1, which was assembled by combining data from
ARTEMIS (687-110 MHz), the  Nan\c{c}ay Decametric Array (DAM, 20-70
MHz), and the WIND/WAVES RAD 2 receiver (7-13.825 MHz; only this
frequency range was relevant to our event). The ARTEMIS radio data
showed no transient emission during the C5.5-class flare.  During the
interval 13:38-13:40 UT, they showed a group of type III bursts
associated with the rise phase of the flare that extends to the
decametric frequency range. Enclosed between two later type III
bursts, there is  a type II burst (marked as ``1" in figure 1) whose
emission drifts from 274 MHz to 154 MHz within 26 s (an enlarged
view of the ARTEMIS dynamic spectrum is given in figure 2). No
evidence of a fundamental-harmonic pair of lanes is available for
this type II burst. We conclude that we observe the harmonic emission
from the shock because in metric type II bursts it is usually stronger
than the fundamental (e.g. Vr\v{s}nak et al. \cite{VAM01}). Continuum emission
is observed in the frequency range 296-570 MHz from 13:40:48 UT to
13:42:12 UT  and also after the type II burst at frequencies from 130
to 398 MHz.

\begin{figure*}
\centering
\includegraphics[width=15cm]{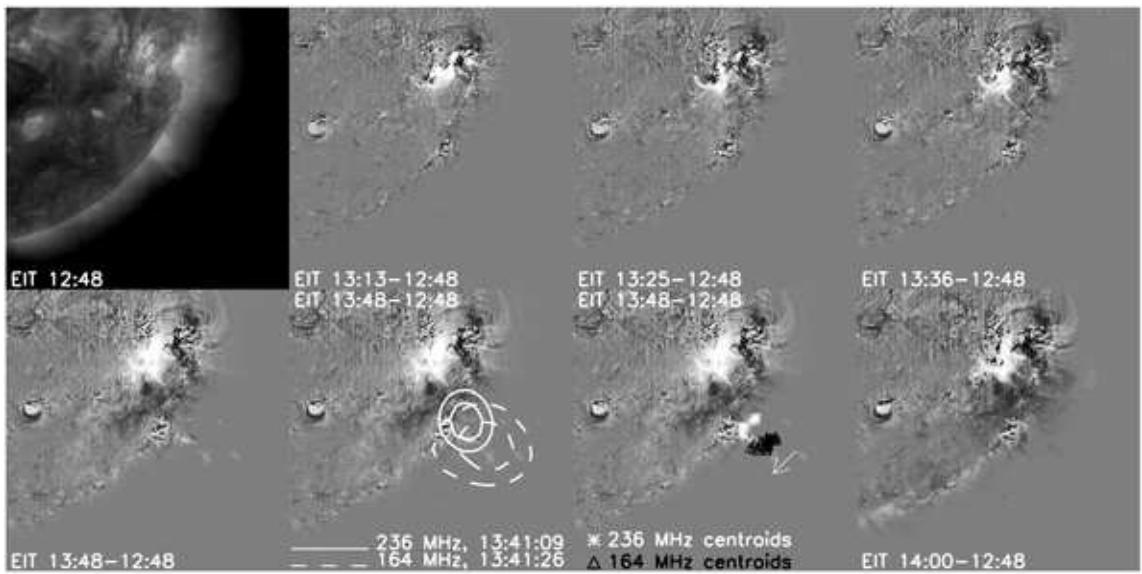}
\caption{EIT 195 \AA\ and NRH data of the 2000 March 2
event. The top-left  panel shows a pre-event image while the other
panels are base difference  images (the pre-event image was subtracted
from each other image). The panel on the right of the left-bottom
panel shows contour plots of the type II source at 236 and 164 MHz
obtained by the NRH at 13:41:09 and  13:41:26 UT, respectively
(contour  levels at 65\% and 85\% of each  frequency's maximum). The
panel on the left of the right-bottom panel shows the centroids of all
NRH  sources associated with the type II burst (white stars for 236
MHz and black triangles for 164 MHz). The arrow denotes the overall
direction of motion of the type II radio source at each frequency.}
\end{figure*}

At lower frequencies, the DAM data display two slowly drifting lanes
of emission (marked F2a and H2a in figure 1), which correspond to the
fundamental and second harmonic emissions of another type II
burst. Unfornately, owing to the gap between the ARTEMIS and DAM
frequencies it is impossible to determine whether the harmonic
emission started at frequencies higher than 70 MHz. The fundamental
emission of the type II burst cannot be clearly distinguished between
13:44  and 13:47 UT in the DAM spectrum, but is probably detected
again in the WAVES spectrum from the highest frequency available to
its receiver (13.825 MHz) down to 7.8 MHz. The harmonic emission of
the type II is  not visible for about 1 minute, but from about
13:45:15 onwards the type II lane marked H2b1 in figure 1  starts and
within less than one minute so does the H2b2 type II lane.  These
lanes show similar frequecy drifts and therefore they may be
considered as ``split bands". The nature of the emission patches
between the F2a and H2b1 lanes is not clear but it is possible that
the fundamental band divided into two lanes.  If we assume that all
the above features are parts of the same shock, its fragmented  nature
can be interpreted in terms of the varying plasma and magnetic field
parameters that suppress radio emission from the shock along a
portion of its trajectory. Finally, it is possible that the features
marked as ``A?" in figure 1 (the slowly drifting emission in the DAM
spectrum  and the blob in the WAVES spectrum) corresponded to an
earlier type II burst. It appeared at 13:30 UT, at the time after the
end of the C5.5-class flare and before the start of the M6.5-class
flare.

The low corona signatures of the event are shown in figure 3. The EIT
data at 13:13 and 13:25 UT show some flare activity in the form of
bright  loops. These brightenings  corresponded to the C5.5 flare
while the M6.5 flare was detected in the remaining images of figure 3.
The difference image at 13:48 UT displays bright small-scale ejecta
emanating from the flare site and from small loops located at the limb
south of the main flare. In the 13:48 and 14:00 UT images,  an EIT
wave can be seen propagating south-southeast from the flare site. Its
computed speed is 410 km/s, in good agreement with the value reported
for the same event by Warmuth et al. (\cite{Warmuth04}).  The dimmings
observed in the EIT data from 13:25 UT onward indicate that the active
region was the source of a CME (images of the CME are given in figure
4).

\begin{figure*}
\centering
\includegraphics[width=15cm]{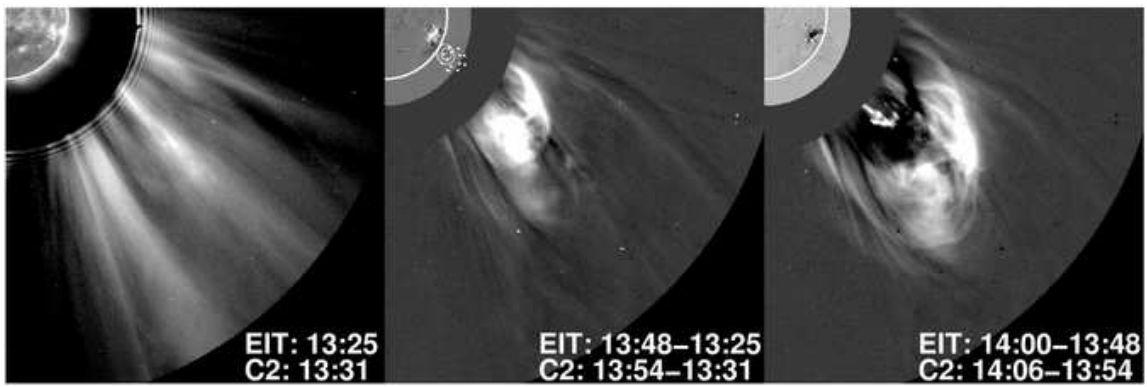}
\caption{Pre-event (left panel) and running difference images (middle and right
panels) of the CME of the 2000 March 2 event observed by the LASCO C2 
coronagraph. In each panel, the LASCO data are combined with the temporally
closest EIT data. In the middle panel, the contours are the same as those
in figure 3.}
\end{figure*}

The range of frequencies of the high-frequency type II burst emission
permitted its imaging at two NRH frequencies, 236 and 164 MHz. Contour
plots of two characteristic maps in each frequency are given in figure
3.  The emission has a compact morphology at both frequencies; the
typical dimensions of the sources were $370\arcsec \times 220\arcsec$
and $270\arcsec \times 170\arcsec$ at 164 and 236 MHz,
respectively. The NRH sources indicate that the shock was moving
almost radially away from the flare site in the general direction of
the ejecta.  In figure 3, the white stars and black triangles denote
the positions of the centroids of the NRH sources associated with the
type II at 236 and 164 MHz, respectively. The arrow denotes that the
overall direction of motion of the radio source at each frequency was
towards south-southeast which is consistent with the motion of the EIT
wave.

The average heights of the centroids of the NRH sources above the
photosphere at 236 and 164 MHz were 0.16$R_{\odot}$ and
0.29$R_{\odot}$, respectively. The most extreme heights were
0.13$R_{\odot}$ and 0.19$R_{\odot}$ at 236 MHz, and 0.25$R_{\odot}$ and
0.31$R_{\odot}$ at 164 MHz. Using the corresponding times, we found that 
the resulting ``average" speed of the shock was 4200 km/s with extreme
values of 3800 km/s (lower limit) and 4400 km/s (upper limit).
This is probably one of the highest coronal shock speeds reported in the
literature (we note, however, that Gergely et al. \cite{Gergely83} reported a
shock speed of 4900 km/s). We note that all reported shock speeds in this
article were not corrected for projection effects to enable us to compare
them directly with the associated CME speeds. However, these corrections
were small and would increase the shock speeds by less than 10\%. 

No imaging data were available for the low-frequency type II burst and
its speed can be estimated from its frequency drift rate if a coronal
density model is provided. The heights of the NRH sources at 236 and
164 MHz were used to select the most appropriate density model. The
method we used is similar to the one presented by Magdaleni\'{c} et
al. (\cite{Magdalenic08}, \cite{Magdalenic10}).  We obtained the
electron densities $N_e$ corresponding to the frequencies
$f=236/2=118$ MHz and $f=164/2=82$ MHz, by employing the equation for
the plasma frequency $f \approx 9 \sqrt{N_e}$, where $N_e$ is
expressed in cm$^{-3}$ and $f$ in kHz (since we assumed harmonic
emission for the type II burst, the NRH frequencies were converted to
the fundamental ones by halving their values). Comparing the radial
dependence of the plasma density behavior for different models, we
found that the 2.8$\times$ Saito model was the most appropriate model
for the ARTEMIS type II burst.  When we considered the extreme heights
of the NRH sources, the densities of the 2.2$\times$ and 3.2$\times$
Saito models were obtained.

If we assume that the 2.8$\times$ Saito density model is also valid
for  the heights accessible to the low-frequency type II burst, we
find (see also figure 5) that  the F2a lane corresponded to heights
from 0.85$R_{\odot}$ to  1.75$R_{\odot}$ above the photosphere and that 
its speed was 710 km/s. The H2a and H2b2 lanes corresponded to heights
from  0.69$R_{\odot}$ to 0.81$R_{\odot}$, and 0.90$R_{\odot}$ to
1.13$R_{\odot}$, respectively, and their derived speeds were 670 km/s
and 560 km/s, respectively. The 2.2$\times$ and 3.2$\times$ Saito
density models yielded heights and speeds that did not vary more than
20\% from the values reported above. However, our method is crude
because:  (1) we applied the density model derived from a previous
type II burst, (2) the CME might have disturbed the densities provided
by any coronal model, and (3) radial propagation of the shock was
assumed whereas no positional information was available.

\begin{figure}
\centering
\includegraphics[width=9cm]{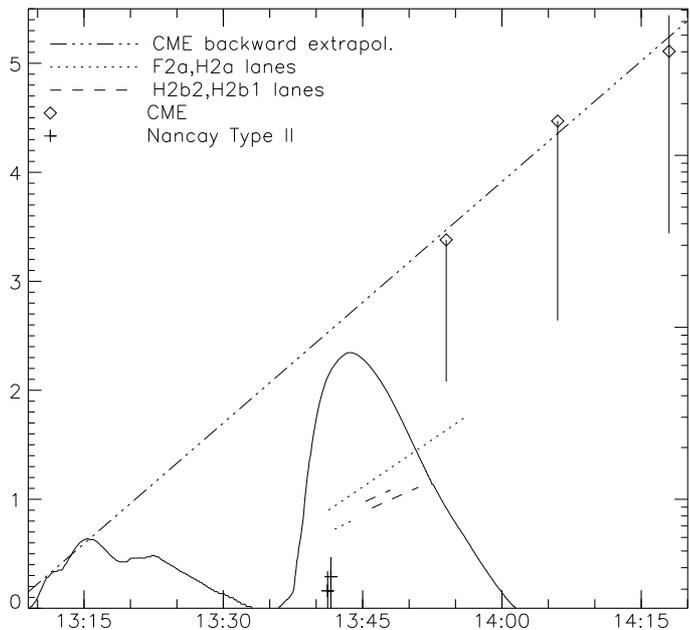}
\caption{Height-time profiles of the various components of the 2000
March 2 event (the vertical axis shows heights above the photosphere expressed
in solar radii). The diamonds denote the heights of the CME front taken
from the CME catalog and the crosses denote the heights of the NRH 
type II sources. The vertical solid lines associated with the diamonds 
denote the height difference between the outermost and innermost  
features of the CME front at a given time. The vertical parts of the crosses 
denote the sizes of the NRH sources. The dashed-dotted curve is the linear
fit to the CME heights. The dotted and dashed curves provide the 
height-time profiles of the F2a-H2a and H2b1-H2b2 lanes, respectively, 
of the low frequency type II using the 2.8$\times$ Saito density model.
The time profile of the GOES X-ray flux in the 1-8 \AA\ channel is also
presented.}
\end{figure}

The CME (see figure 4) was present in the LASCO C2 field of view at
13:54 UT and 14:06 UT, while its first appearance in the C3 field of
view occurred at 14:18 UT. The height of the CME was measured at
several fixed position angles along its leading front and the results
appear in figure 5. In the same figure, the diamonds denote the
heights reported in the LASCO
catalog\footnote{http://cdaw.gsfc.nasa.gov/cme\_list}.  In most cases,
these heights correspond to the outermost CME front propagation. The
CME speed in the time interval 13:54-14:06 UT was between 800 and 1050
km/s, while in the time interval 14:06-14:18 UT it was between 780 and
1010 km/s. Using all the C2 and C3 images in which the CME was
present, we found that its average speed was between 750 and 880 km/s
with a speed of 850 km/s corresponding to the heights along the
position angle that intersects the outermost front features.    For
the computation of the lateral expansion of the CME, we measured the
position angles of the intersections of its northern and southern
flanks and the solar limb, which were then converted to the
along-the-limb distances from the flare site. We found that the
lateral speed of the southern and northern flank was 260 and 150
km~s$^{-1}$, respectively.

The ability of mass transients to drive shocks depends on the
relationship of their speed to the local Alfv\'{e}n speed.  The band
split of type II emissions can be attributed to simultaneous emission
from the plasma ahead and behind the shock (Smerd et al.
\cite{Smerd75}; Vr\v{s}nak et al. \cite{VAM01}) and can be used to
estimate the Alfv\'{e}n Mach number and the Alfv\'{e}n speed. Under
the quasi-perpendicular shock approximation and plasma beta $\beta=0$,
the Alfv\'{e}n Mach number $M_A$ is related to the compression $X$ as

\begin{equation}
M_A = \sqrt{\frac{X(X+5)}{2(4-X)}} \,\,,
\end{equation}  
where $X=(1 + BDW)^2$ and $BDW$ is the average relative frequency
separation between the bands (see Priest \cite{Priest82}; Vr\v{s}nak
et al. \cite{Vrsnak02}).  We measured the central frequencies  $f_1$
and $f_2$ of the split bands of the harmonic emission of the low
frequency type II burst (lanes H2b1 and H2b2 of figure 1)  during
several time  intervals and found that their $BDW=(f_2-f_1)/f_1$ was
0.31. From equation (1), then we obtained $M_A=1.59$. For a shock
speed of $v_{shock}=560$ km/s (see above), the resulting Alfv\'{e}n
speed was $v_A=v_{shock}/M_A=350$ km/s, which is within the values
reported in the literature (e.g. Vr\v{s}nak et al. \cite{Vrsnak02}).

The speed of the CME exceeded the estimated Alfv\'{e}n speed but the CME 
is unable to drive the high frequency shock. The primary argument is the
discrepancy of a factor of 4-5.25 between its own speed and the CME
speed, if we use the average shock speed of 4200 km/s and the first
measurements of the CME speed. Using the lower and upper limits of the
shock speed, we find factors of 3.6-4.75 and 4.2-5.5, respectively, for
the discrepancy between the shock speed and the first measurement of the
CME speed.

The early evolution of the CME cannot be determined because of the
size of the C2 occulting disk. However, if it exhibited an
accelaration phase before entering the LASCO field of view, it is
unlikely that it obtained a  speed of about 4000 km/s. If that were
the case, after the acceleration the CME would have to decelarate from
about 4000 km/s to about 1000 km/s  within 12 minutes (this is the
time from the end of the high frequency type II burst until the  first
appearance of the  CME in the LASCO images). That would lead to a
deceleration of -4.17 km/s$^{-2}$, which is implausible. It is well
established that the speed of CMEs  does not change much after they
obtain their terminal speed by the end of the  acceleration phase; a
typical CME deceleration is about -20 m/s$^{-2}$ (Gopalswamy et
al. \cite{Gopalswamy01}), and only in exceptional cases can reach
values of about -150 m/s$^{-2}$ (Vr\v{s}nak \cite{Vrsnak01b}).

The high-frequency type II burst could have been driven by a CME only
if (1)  the CME was so fast that it could not be observed by the C2
coronagraph, which would then have insufficient cadence, and (2) its
signal was so weak that it could not be imaged by the C3
coronagraph. The first requirement implies that the CME would need to
propagate from the height of the NRH type II sources    to the edge of
the C2 field of view with an average speed of about 4550 km/s, which
is consistent with the speed of the shock.

Under the assumptions of radial propagation and a 2.8$\times$ Saito
density model, we estimated by extrapolation to the photospheric
height the onset time of the type II bursts using their speeds and the
heights corresponding to their appearances in the dynamic spectra.
The backward extrapolation of the type II burst observed by ARTEMIS
and  NRH yields its onset time at 13:40:03 UT, which lies within the
rise phase of the M6.5-class flare. This may indicate that the shock
could be generated by the explosive energy release in the flare (see
also Pothitakis et al. \cite{Pothitakis09}). From backward
extrapolation of the observed CME, we conclude that it could be
related to the earlier C5.5-class flare (see figure 5) and  this is a
second argument against the assumption that the high-frequency type II
burst was ignited by that CME.

The situation is less clear concerning the origin of the shock
observed by DAM and WAVES. The backward extrapolations (see figure 5)
of the type II lanes associated with this shock provide onset times of
between 13:20:25 and 13:25:27 UT. This time interval lies around the
secondary peak of the C5.5-class flare that occurred before the
M-class flare (see figure 5).  We note that the backward extrapolation
of the ``A?" lane of figure 1 yields its onset time at 13:14:30 UT,
which lies within the rise phase of the C5.5-class flare.

Figure 5 indicates that the low frequency shock was 0.35$R_{\odot}$
and 1.6$R_{\odot}$ behind the innermost and outermost features,
respectively, of the  CME front when the CME first appeared in the
LASCO C2 field of view. This  result is sensitive  to the density
models used for the calculation of the height-time curves of the
shock. The shock height would match the height of the innermost and
outermost features of the CME front if the densities of the
2.8$\times$ Saito model decreased by a factor of 1.5 and 5.4,
respectively. Furthermore, figure 4 indicates that at 13:54 UT the
streamer south of the CME was deflected. This may be evidence of a
CME-driven shock running ahead of the CME (e.g. Sheeley et
al. \cite{Sheeley00}; Vourlidas et al. \cite{Vourlidas03}; Ontiveros
and Vourlidas \cite{Ontiveros09}; Liu et al. \cite{Liu09}).  This in
turn may mean that the shock did not move radially outward because of
its refraction into regions of low Alfv\'{e}n speeds (e.g.  Uchida et
al. \cite{Uchida73}). In this case, the calculation of the speed of
the shock using the frequency-drift rate of the type II provides only
a lower limit to the true speed of the shock (e.g. Pohjolainen et al.
\cite{Pohjolainen07}).

\begin{figure}[h]
\centering
\includegraphics[width=9cm]{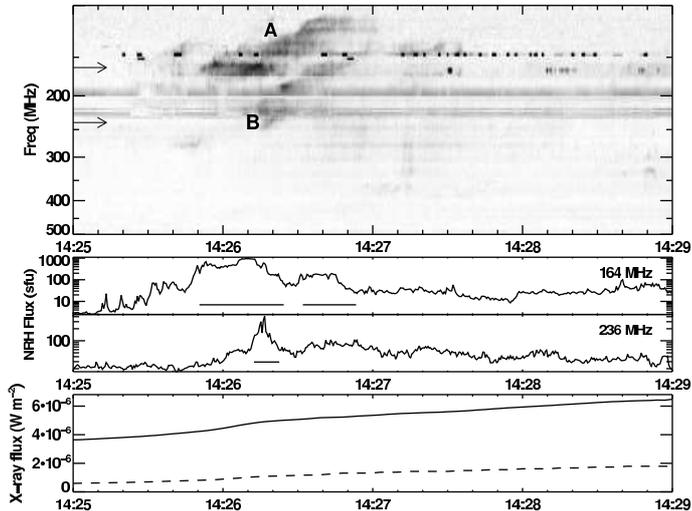}
\caption{Same as in figure 2 for the 2000 March 7 event with the
exception that there is an additional panel where the time profiles of the 
GOES X-ray flux in the 1-8 \AA\ (solid line) and 0.5-4 \AA\ (dashed line)  
energy channels are presented.}
\end{figure}

\subsection{The 2000 March 7 event}

The type II burst of 2000 March 7 was associated with a GOES C6.3
class flare that occurred in AR NOAA 8891 (13\degr S 60\degr W). The
flare started at 14:21  UT and reached its maximum at 14:30 UT.  The
dynamic spectrum of the event from ARTEMIS is presented in figure 6
(see also Table 2). Before the type II burst, there were no signatures
of  transient activity in the ARTEMIS spectral records but the WAVES
data showed at least four interplanetary type III bursts from about
14:24:15 until about 14:29:30 UT (no DAM observations were available).
The NRH 10-s resolution data, however, detected a
compact source above the active region at 432 and 164 MHz that was
probably associated with a noise storm.  According to the catalog of
type II bursts compiled by the Astrophysical Institute of Potsdam
(AIP), the type II burst of figure 6 showed both fundamental and
harmonic emission bands; the observing frequencies of ARTEMIS provided
data for the harmonic emission band only, which was split into two
well-defined lanes marked A and  B in figure 6. Their relative
frequency separation $BDW$ (see section 3.1)  varied from 0.30 to
0.33.   After the type II burst and until about 14:50 UT, the ARTEMIS data
showed continuum emission from 300-400 to 130 MHz. Another type II
burst was embedded into the continuum, and displayed two patchy emission
lanes that drifted from 315 MHz at 14:30:52 UT to 185 MHz at 14:32:33
UT.

Using the method described in section 3.1, we found that the backward
extrapolation  of lanes B and A, yielded the onset time of the  type
II burst at 14:24:12 and 14:24:50 UT, respectively. We note that the
results of this method are practically independent of the coronal
density model used to calculate the height and speed of the radio
source because a higher density model yields larger heights but also
higher speeds and the two effects cancel out (Vr\v{s}nak et al. 
\cite{Vrsnak95}).

\begin{figure*}
\sidecaption
\includegraphics[width=11cm]{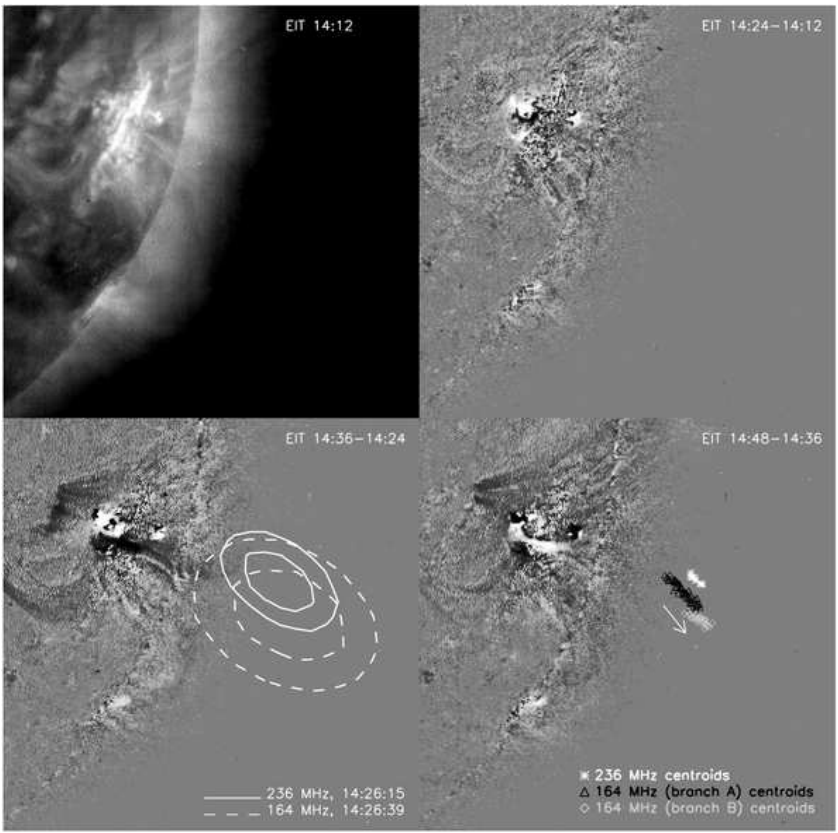}
\caption{EIT 195 \AA\ and NRH data of the 2000 March 7 event in the
same format as in figure 3, with the following exceptions: All EIT
images except the one of the top-left panel are running difference
images.  The bottom-left panel shows contour plots of the type II
source at 236 and 164 MHz obtained by the NRH at 14:26:15 and 14:26:39
UT,  respectively. The right-bottom panel shows the centroids of all
NRH sources associated with the type  II burst (white stars for 236 MHz,
black triangles for 164 MHz sources of branch A, and gray diamonds
for 164 MHz sources of branch B).}
\end{figure*}

The EIT data for the event are presented in figure 7.  The flare
appears as rather compact brightenings in the EIT images. At 14:36 UT,
there was evidence of some activation of a filament-like elongated
dark  feature whose projection on the solar disk ran almost
perpendicularly to the bright features of the active region (the
activation probably started earlier as indicated by the dark
semi-circular feature south of the most prominent brightening of the
14:24 UT difference image).  In the two bottom panels of figure 7,
diffuse loop-like features without appreciable motion appear on the
limb south of the flare brightening.

The type II burst was imaged by the NRH at 236 and 164 MHz. In figure
7, we present contour plots of two characteristic maps in each
frequency as well as the positions of the centroids of the NRH sources
associated with the type II burst. The emission was unresolved at both
frequencies and the typical dimensions of the sources were $150\arcsec
\times 290\arcsec$ and $260\arcsec \times 430\arcsec$ at 236 and 164
MHz, respectively. The alignment of the type II source positions
deviates from being radial with respect to the associated flare. The
non-radial propagation of type II sources has been reported in other
cases as well (e.g. Nelson and Robinson \cite{Nelson75}; Klein et
al. \cite{Klein99}). Three possible interpretations have been
considered for such propagation: refraction of the  shock wave into
regions of low Alfv\'{e}n speed (Uchida et al. \cite{Uchida73}),
preferential acceleration of electrons in restricted regions of the
shock front  (e.g. Steinolfson \cite{Steinolfson84}), or suitable
orientation of the motion of the shock driver as well as the magnetic
environment where the type II burst occurs (Aurass et al. \cite{Aurass98}; 
Klein et al. \cite{Klein99}).

\begin{figure*}
\sidecaption
\includegraphics[width=11cm]{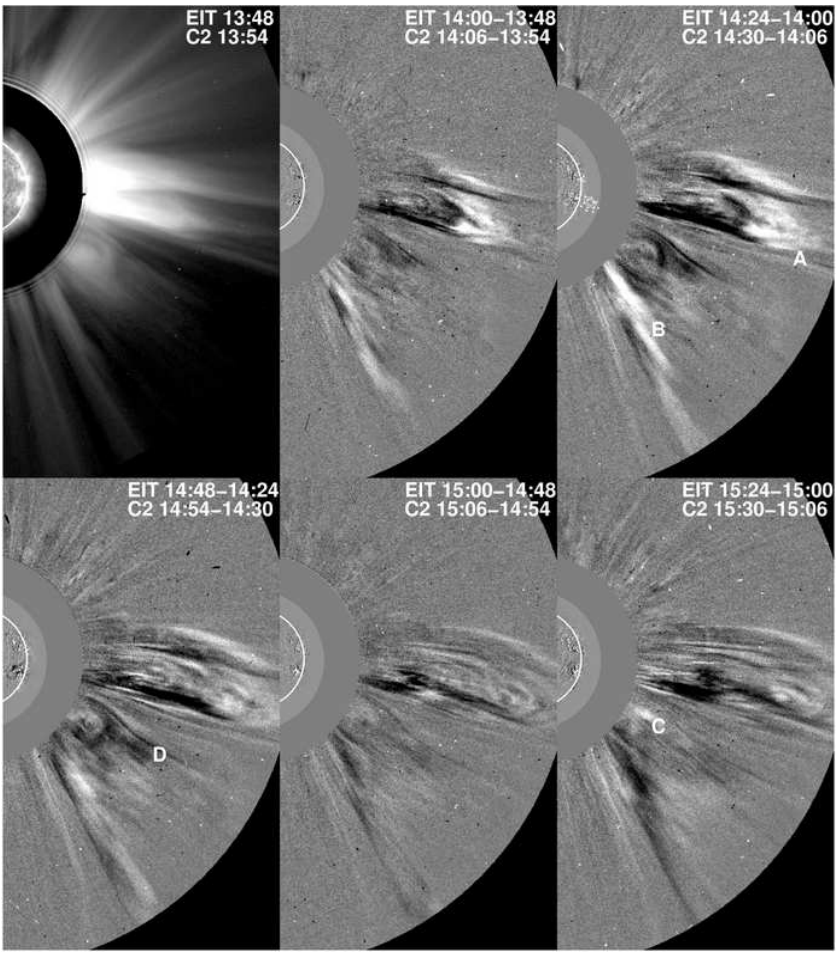}
\caption{Pre-event (top-left panel) and running difference images of the 
2000 March 7 event observed by the LASCO C2 
coronagraph. In each panel, the LASCO data are combined with the temporally
closest EIT data. In the top-right panel, the contours are the same as the ones
in figure 7.}
\end{figure*}

The NRH observations imaged both type II bands at 164 MHz.  The sizes
of the sources were similar for both bands and their locations did not
overlap; overall, the high frequency band (B in figure 6) was further
from the solar surface than the low frequency band (A in figure 6). In
the bottom-right panel of figure 7, the arrow indicates the general
direction  of motion of the radio source at each frequency, which is
consistent with the ARTEMIS data indicating that the 164 MHz emission
of the low frequency branch appeared earlier than that of the high
frequency band. Imaging observations of split-band sources with modern
instruments is not frequent; the old Culgoora Radioheliograph data
(e.g. Smerd et al.  \cite{Smerd75}; Nelson and Robinson
\cite{Nelson75}) showed the split-band sources to be cospatial on
average; when they were not, either band source could be further from
the Sun.

The average positions of the centroids of the NRH sources at 236 MHz
and those of the high-frequency band at 164 MHz yielded a shock speed
of 2300 km/s. If the most extreme positions and times are considered,
we obtain a lower limit of 2100 km/s and an upper limit of 2400
km/s for the speed of the shock. Given the non-radial propagation of
the shock, the use of the drift rate from the spectral data could
provide misleading results for the speed of the shock.

In figure 8 we present images from the LASCO C2 coronagraph before and
after the type II burst. There were two CMEs in the field of view of
the instrument (marked as A and B in the figure) but none of them were
related to  our event, as becomes clear in figure 9. Linear fits
to their height-time points presented in figure 9 yielded onset times at 12:17
and 13:02 UT for CMEs A and B, respectively. Furthermore, the timing
of the dark feature marked as D in figure 8 rules out its possible
contribution to the shock formation. This feature was probably the
remnant of a previous CME, which moved with a speed of 230 km/s
between 14:30 and 14:54 UT and then disappeared.  In the 15:30 C2
image, there is a  narrow bright feature (marked as C in figure 8) in
the general  direction of the prolongation of the line that connects
the flare site and  the 236 MHz sources. This feature did not appear
in the subsequent C2 image, which  was taken at 15:54 UT.  Under the
assumption that its speed was constant from the height of the 236 MHz
source to its height in the C2 image, its estimated speed was 260 km/s. 
Our data do not allow us to evaluate its possible contribution to the
shock formation in detail. However, it is not likely to be associated
with the dark feature whose activation was observed in the EIT images
because after its activation it showed no outward motion in the EIT
field of view.

\subsection{The 2000 May 2 event}

The type II burst of 2000 May 2 was associated with a GOES M2.8-class
flare that occurred at 22\degr N 68\degr W (NOAA AR8971) and a CME.
The dynamic spectrum of the event from ARTEMIS is presented in figure
10.  In the ARTEMIS frequency range, the event started with three
groups of type III bursts associated with the rise phase of the
flare. The WAVES data showed that these bursts extended into the
interplanetary medium (no DAM observations were available). The type
II burst  appeared toward the end of the metric type III activity at
about 14:48:04 UT and lasted until 14:52:53. For two minutes, it
drifted from 340 MHz to 221 MHz. No trace of it could be found in the
ARTEMIS spectrum from 14:50:05 to 14:51:24 UT. As mentioned in section
3.1, these interuptions may reflect varying plasma and magnetic field
parameters which suppress radio emission from the shock along a
portion of its trajectory.  The type II appeared again at 170 MHz,
then drifted to 127 MHz within 90 s. During that time interval, a
short  continuum was present in the dynamic spectrum close to the type
II burst. There was no trace of a fundamental-harmonic pair  of lanes
for this type II in either the ARTEMIS spectrum or in the available
catalogs of type II bursts (the burst appears only in the NOAA catalog
as observed from the Bleien station at frequencies from 220 to 280
MHz). As in the case of the 2000 March 2 ARTEMIS type II burst, we
attributed the observed lane to harmonic emission.

\begin{figure}
\centering
\includegraphics[width=8cm]{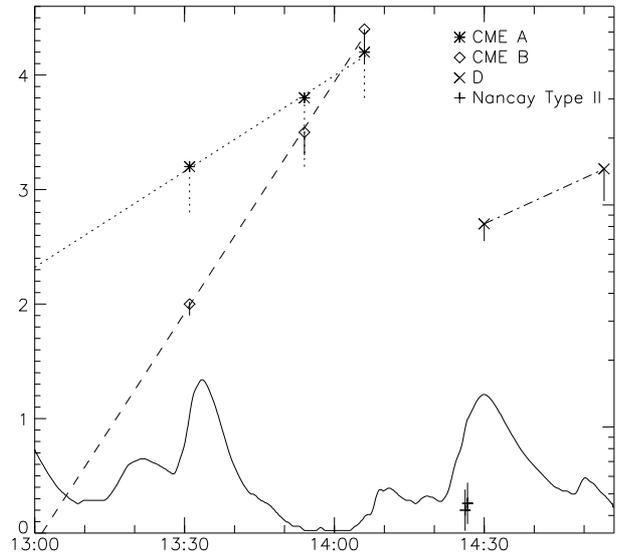}
\caption{Height-time profiles of various features of the 2000 March 7 
event in the same format as in figure 5,
with the  following exceptions: The stars, diamonds and ``x"s denote
the heights of features A, B, and D of figure 8.  The crosses
denote the heights of the NRH type II sources of branch B. The
vertical solid and dotted lines associated with the diamonds,
stars, and ``x"s denote height difference between the outermost
and innermost  features of their fronts at a given time.}
\end{figure}

The EIT data for the event are presented in figure 11. The images taken
at 14:36 and 14:48 UT did not show any prominent brightenings, but
did show expanding loop-like features. The height of the tip of
the  expanding loops above the solar surface was 61\arcsec\ and
91\arcsec\ at 14:36 and 14:48 UT, respectively, yielding a speed of
about 30 km/s.  In the 14:36 UT image, there were two dark patches
below the activated loops. The dark areas grew in the two subsequent
images and probably reflected the dimming of the EUV emission caused
by the mass depleted by the CME. The last EIT image was obtained after
the end of the type II burst and shows an intense brightening
associated with the flare. It also shows large-scale dark loop-like
features, which probably represent loops that were carried away by the
CME.

\begin{figure}[h]
\centering
\includegraphics[width=9cm]{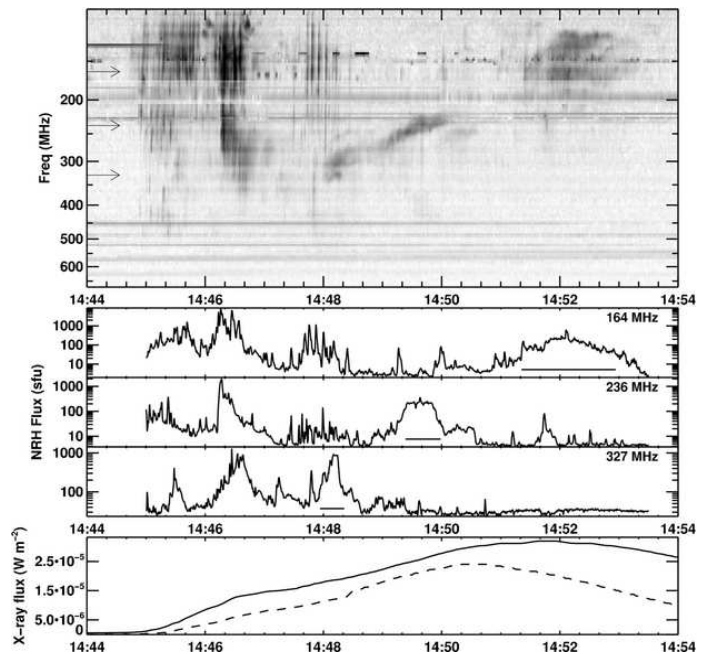}
\caption{Same as in figure 6 for the 2000 May 2 event.}
\end{figure}

Before the event, the NRH 10-s resolution data at 432 MHz detected a
compact source at the limb, probably associated with the active region
and diffuse emission bridging that source with another source located
at the southwest quadrant just above the limb. At 164 MHz, the
pre-event data showed a compact source located about 0.7$R_{\odot}$
beyond the limb in the southwest quadrant of the image.

The type II burst was imaged by the NRH at 327, 236, and 164
MHz. Contour plots of one characteristic map at each frequency as well
as the positions  of the centroids of the NRH sources associated with
the type II are given in figure 12. The emission of the type II burst
consisted of one source at 327 and 164 MHz with typical sizes of
$150\arcsec \times 210\arcsec$ and  $340\arcsec \times 220\arcsec$,
respectively. The 236 MHz emission consisted of two sources; the
stronger was located radially below the 164 MHz sources and the weaker
was located radially above the 327 MHz sources. This may indicate that
inhomogeneities in the local plasma density and/or magnetic field
prevented the formation of radio emission from the entire shock front
at 327 and 164 MHz.  Overall, the alignment of the centroids of
the 327 and 236 MHz emissions was non-radial with respect to the flare
site, in contrast to the overall alignment between the  236 and 164
MHz centroids. With respect to the expanding loops, the 327 MHz
sources appeared at the side of their northernmost features, while the
centroids of the 236 and 164 MHz appeared above them.

The shock speeds derived from the average positions of the centroids
of the 327 and 236 MHz sources and  from those of
the  centroids of the 236 and 164 MHz  sources were 1240 and 1180
km/s, respectively. If the most extreme positions and times are
considered, we obtain a lower limit of 1030 km/s and an upper limit
of 1860 km/s for the former speed and a lower limit of 930 km/s
and an upper limit of 1440 km/s for the latter speed.

\begin{figure*}
\centering
\includegraphics[width=11cm]{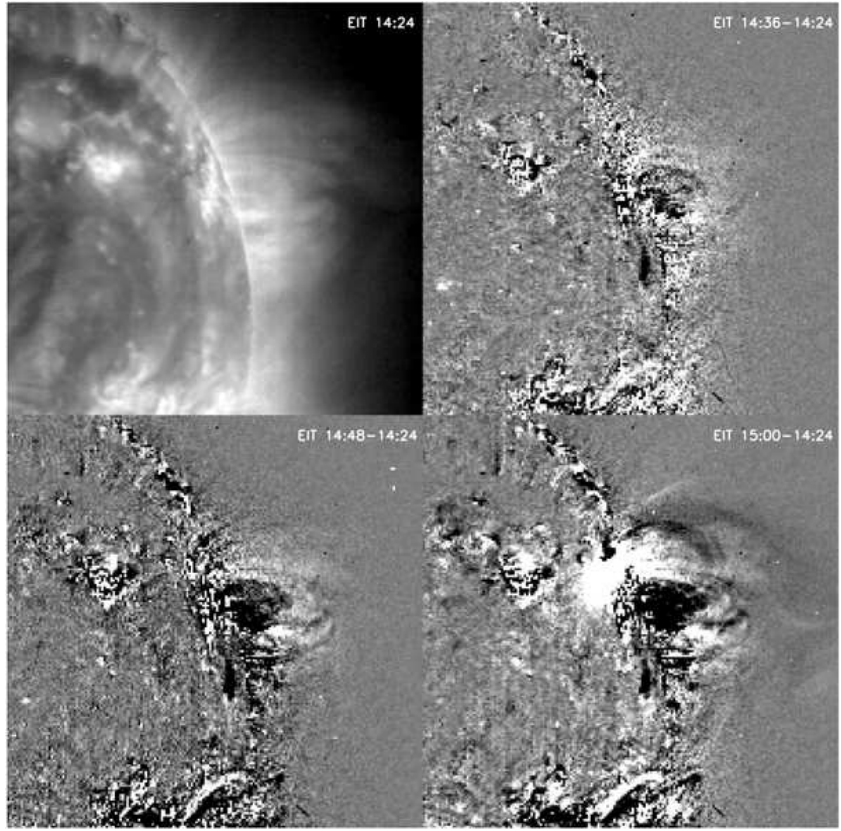}
\caption{EIT 195 \AA\ data of the 2000 May 2 event. The top-left panel shows
a pre-event image, while the other panels show base difference images.}
\end{figure*}

\begin{figure*}
\centering
\includegraphics[width=11cm]{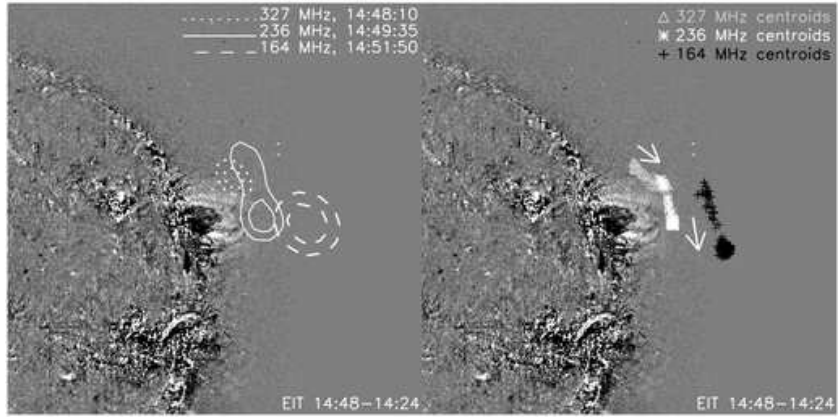}
\caption{The left panel shows contour plots of the 2000 May 2 type II
source  at 327, 236, and 164 MHz obtained by the NRH at 14:48:10,
14:49:35, and  14:51:50 UT, respectively (contour levels at 65\% and
85\% of each frequency's maximum). The right panel shows the centroids
of all NRH sources associated with the type II burst (gray triangles
for 327 MHz, white stars for 236 MHz, and black crosses for 164
MHz). The background image is an EIT difference image. The upper arrow
denotes the overall direction of motion of the type II source at 327
MHz and the lower arrow denotes the overall direction of motion of the
type II source at 236 and 164 MHz.}
\end{figure*}

Two CMEs appear in the difference images of figure 13; the one
associated with our event was the wider one  (the north CME). The
height-time profile of the CME is given in figure 14. For the
subsequent computations, all heights were measured along the line
connecting the disk center with the outermost feature of the CME front at
15:06 UT (in figure 14, the resulting CME heights correspond to the 
diamonds). The position angle of this line
measured from Solar west counter-clockwise was 19\degr.  The backward
extrapolation of the CME heights using all LASCO images in which the CME
appeared  (i.e. not only the three images that are represented in
figure 14) yielded the  CME onset time at 14:40 UT. This computation
is crude because it does not take into account the possible
acceleration of the CME when it was below the C2 occulting disk. We
tried to address this problem by considering the expanding EIT loops
at 14:36 and 14:48 UT as signatures of the CME in its infancy. Their
projected heights are marked  with the squares of figure 14. The
projected CME heights (including the squares of figure 14) were fitted
with the function

\begin{eqnarray}
H(t) &=& H(t_1) + \frac{1}{2} (v_f+v_0)(t-t_1) \nonumber \\
&+& \frac{1}{2} (v_f-v_0) \tau \ln \left[ \cosh \left( \frac{t-t_1}{\tau} \right) \right] \,\,,
\end{eqnarray}
where $v_0$ and $v_f$ are the initial and final CME speeds,
respectively, $t_1$ is the time of peak acceleration, and $\tau$ is
the timescale of the rise to peak acceleration. This function was
used to fit CME height-time profiles by Sheeley et al. (\cite{Sheeley07}) 
and Patsourakos et al. (\cite{Patsourakos10}) and is capable  of 
reproducing a wide range of profiles from almost constant acceleration 
to impulsive acceleration.

Our best-fit model (the dashed curve of figure 14) showed that the CME
acceleration peaked at 14:48:15 UT, at a height $H(t_1)=0.2R_{\odot}$
and  lasted for 215 s. It is interesting that the peak acceleration
occurred 11 s after the start of the type II burst. The values of
$v_0$ and $v_f$ used in our model were  27 km/s and 1400 km/s,
respectively. The maximum value of the acceleration of the CME was 3.2
km/s$^{-2}$. At the times when the 236 and 164 MHz shock emission occurred,
the model yielded CME speeds of about 1130 and 1380 km/s,
respectively. These speeds were derived under the assumption of radial
CME propagation whereas the alignment of the centroids of the 327 and
236 MHz sources deviated significantly from the radial direction. The
speed of the shock derived from the positions of the 327 and 236 MHz
sources had a component along the assumed CME direction of propagation
of about 740 km/s.  Therefore, the kinematics of the CME and the
shock were broadly  consistent.

\section{Discussion and conclusions}

\begin{figure*}
\centering
\includegraphics[width=11cm]{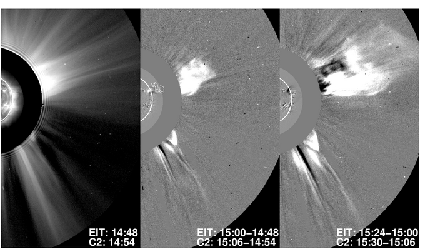}
\caption{Same as in figure 4 for the 2000 May 2 event. In the middle panel,
the contours are the same as the ones in figure 12.}
\end{figure*}

We have studied three events associated with metric type II bursts.
The purpose of our study was to determine the origins of
the shocks. To take advantage as much as possible of projection
effects, we selected events that occurred away from disk center
(longitudes equal or larger than 60\degr). The ambiguity about the
location of the shock radio sources was partially removed by analyzing
events for which the type II emission was imaged at least at two
frequencies by the NRH.

In all events, both flares and CMEs were observed and their
contributions to the initiation of the shocks were discussed in
section 3. The onset times of all type II bursts observed by ARTEMIS 
were estimated from the backward extrapolation of their emission lanes and 
were found to lie between the flare onset and the flare peak. However, this 
is a weak argument in favor of the flare-ignited-shock scenario in the
cases where there is a tight temporal relationship between the flare
and the CME as can be clearly seen for the 2000 May 2 event. We excluded
the CMEs as primary drivers of the high frequency shock of 2000 March
2 and of the 2000 March 7 shock using additional arguments.

The high frequency 2000 March 2 shock is one of the fastest metric
type II bursts reported in the literature. The NRH sources at 236 and
164 MHz were aligned radially with respect to the flare site and
allowed us to derive a speed of about 4200 km/s without using the
drift rate or any coronal  density model. On the other hand, the speed
of the CME that was first observed 12.5 minutes after the end of the type
II burst  was about 850 km/s. The backward extrapolation of the CME
showed that it was probably related to a smaller flare that occurred
prior to the one linked to the metric type II onset. The CME was
observed by the LASCO C2 and C3 coronagraphs and its early evolution
was missed.  However, it is unlikely that it attained a speed of about
4000 km/s during its acceleration phase because when the
acceleration phase ceases, CMEs propagate with a more or less constant
speed that is similar to the speed they had acquired by the end of their
acceleration (e.g. Zhang et al.  \cite{Zhang01}).

That two type II bursts were observed in close time succession during
the 2000 March 2 event (i.e. the one observed by ARTEMIS and NRH and
the one observed by DAM and WAVES) provides an extra argument that
both of them could not have been ignited by CMEs because the LASCO
observations did not show evidence of two different successive
CMEs. Cases of successive type II bursts had been reported previously
(e.g. Shanmugaraju  et al. \cite{Shanmugaraju05}; Subramanian and
Ebenezer \cite{Subramanian06}; Magdaleni\'{c} et
al. \cite{Magdalenic08}). In the absence of observations of
successsive CMEs, the initiation of these shocks is interpreted in
terms of successive bursts of energy release in the flare or in terms
of a CME driver for one of them and an ignition by the flare energy
release for the other.

The origin of the low frequency type II burst of the 2000 March 2
event remains unclear. No imaging observations were available at
decametric frequencies and thus its trajectory was estimated from its
drift rate and the best-fit coronal density model derived from the
positions of the sources of the high frequency type II burst. The
resulting height of the shock radio source was 0.35$R_{\odot}$ behind
the innermost feature of the CME front when the CME first appeared in
the C2 field of view. However, this is not a strong indication of a
blast wave origin of the shock because the deflection of a streamer
south of the CME was observed. The streamer deflection might have been
caused by a shock running ahead of the CME (e.g.  Liu et
al. \cite{Liu09} for a recent example). If that were the case, the
propagation of the shock would not be radial and the speed calculated
from its drift rate would provide a lower limit to its true speed.

The 2000 March 7 type II burst was the second shock of our study for
which a  CME driver seemed inappropriate. The shock speed was about
2300 km/s and was estimated from the positions of the NRH sources at
both 236 and 164 MHz.  The CMEs observed by the LASCO before and after
the event were not related to the shock; from the available LASCO
data, their onset times were estimated 2.15 and 1.4 hours before the
appearance of the type II burst. The rise of the CMEs  was not
synchronized with the rise phase of the flare, while the onset time of
the type II burst fell between the onset and the peak energy release
of the flare. Similar cases were reported  by Magdaleni\'{c} et
al. (\cite{Magdalenic08}; \cite{Magdalenic10}).

\begin{figure}
\centering
\includegraphics[width=9cm]{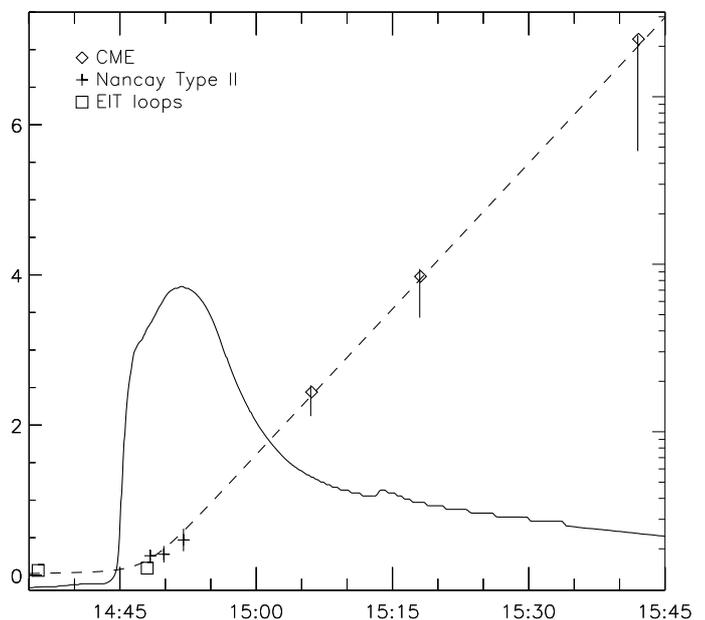}
\caption{Height-time profiles of various components of the 2000 May 2
event in the same format as in figure 5, with the following
exceptions. The squares  mark the heights of the expanding loops seen
in EIT images and the dashed line marks the best fit of the CME
heights by the function given in eq. (2) (see text for details).}
\end{figure}

The 2000 May 2 type II burst was imaged at three frequencies by the
NRH. The trajectory of the shock was complicated: the centroids of the
236 and 164 MHz sources were aligned radially with respect to the
flare site but the 327 MHz sources were displaced laterally.  Possible
interpretations were discussed in section 3.2 (the sources of the 2000
March 7 type II burst also showed non-radial alignment with respect to
the flare site). The EIT images showed expanding loops that were
probably the early signatures of the CME.  The acceleration phase of
the CME occurred before its appearance in the C2 field of
view. However, we were able to reconstruct it by fitting the heights
of the EIT expanding loops and the heights of the white-light CME
front to the same function. The acceleration peaked around the time of
the appearance of the type II burst, at a height of 0.2$R_{\odot}$ and
lasted 215 s. Previous studies have shown that the ejected plasma can
be accelerated within a couple of minutes and at heights as low as
0.05$R_{\odot}$ (e.g. Dere et al. \cite{Dere99}; Zhang et
al. \cite{Zhang01}; Gallagher et al.  \cite{Gallagher03}; Williams et
al. \cite{Williams05}). The kinematics of the CME were broadly
consistent with the kinematics of the shock sources implying that the
CME is the most likely cause for  the shock ignition.

That two of the type II bursts we studied were not driven by CMEs does
not necessarily imply that there was no material driver at all; it is
possible that an initial driver (which does not develop into a
large-scale CME) of coronal shocks is at work even if they later
become blast waves.  Unfortunately, the cadence of our EUV data was
insufficient to test whether this was the case for the small-scale
ejecta observed in the 2000 March 2 event and the dark feature that
was activated during the 2000 March 7 event.

Our analysis illustrates the diversity of the physical conditions
that may lead to the generation of coronal shocks:

\begin{enumerate}

\item The high speed coronal shock wave (2000 March 2) was
associated  with an M6.5-class  flare. A CME that was observed after
the end of the metric type II burst  had a speed of a factor of 4-5.25
smaller than the speed of the shock. Hence, the shock could not have
been ignited by the CME. That CME may have contributed to the ignition
of a decametric shock that appeared just after the  high speed metric
shock and was related to an earlier C5.5-class flare.

\item The shock wave observed on 2000 March 7 was synchronized with the 
rise phase of the flare but not with the CMEs that appeared in the
coronagraph data.

\item The last event (2000 May 2) included a shock wave whose
kinematics  were broadly consistent with the kinematics of the
associated CME inferred from EUV and coronagraph data. Thus, the CME
is the primary candidate for the ignition of the shock.

\end{enumerate}

\begin{acknowledgements}
We thank A. Kerdraon and K.-L. Klein for providing the NRH visibities
and the DAM data, respectively. We also thank S. Patsourakos for useful
discussions.
\end{acknowledgements}


\begin{thebibliography}{}

\bibitem[1998]{Aurass98} Aurass, H., Hofmann, A., Urbarz, H.-W. 1998, A\&A,
334, 289

\bibitem[2005]{Cane05} Cane, H.V., \& Erickson, W.C. 2005, ApJ, 623, 1180

\bibitem[2001]{Caroubalos01} Caroubalos, C., Maroulis, D., Patavalis, N., et
al. 2001, Exp. Astron., 11, 23

\bibitem[2004]{Caroubalos04} Caroubalos, C., Hillaris, A., Bouratzis, C., et
al. 2004, A\&A, 413, 1125

\bibitem[2005]{Cho05} Cho, K.S., Moon, Y.J., Dryer, M., et al. 2005, JGR,
110, A1201

\bibitem[2007]{Cho07} Cho, K.S., Lee, J., Moon, Y.J., et al.
2007, A\&A, 461, 1121

\bibitem[2005]{Ciaravella05} Ciaravella, A., Raymond, J.C., Kahler, S.W.,
Vourlidas, A., \& Li, J. 2005, ApJ, 621, 1121

\bibitem[2002]{Classen02} Classen, H.T., \& Aurass, H. 2002, A\&A, 384, 1098

\bibitem[2006]{Dauphin06} Dauphin, C., Vilmer, N., \& Krucker, S. 2006,
A\&A, 455, 339

\bibitem[1999]{Dere99} Dere, K.P., Brueckner, G.E., Howard, R.A., Michels, D.J.,
\& Delaboudiniere, J.P. 1999, ApJ, 516, 465

\bibitem[2003]{Gallagher03} Gallagher, P.T., Lawrence, G.R., Dennis, B.R. 2003,
ApJ, 588, L53

\bibitem[1984]{Gary84} Gary, D.E., Dulk, G.A., House, L., Illing, R., et al.
1984, A\&A, 134, 222

\bibitem[1983]{Gergely83} Gergely, T.E., Kundu, M.R., \& Hildner, E. 1983,
ApJ, 268, 403

\bibitem[2006]{Gopalswamy06} Gopalswamy, N. 2006, American Geophysical Union
Geophysical Monograph Series, 165, 207

\bibitem[1998]{Gopalswamy98} Gopalswamy, N., Kaiser, M.L., Lepping, R.P., 
et al. 1998, JGR, 103, 307

\bibitem[2001]{Gopalswamy01} Gopalswamy, N., Yashiro, S., Kaiser, M.L., Howard,
R.A., \& Bougeret, J.-L. 2001, JGR, 106, 29219

\bibitem[2005]{Gopalswamy05} Gopalswamy, N., Aguilar-Rodriguez, E., Yashiro, S.,
et al.  2005, JGR, 110, A12S07

\bibitem[2009]{Gopalswamy09} Gopalswamy, N., Thompson, W.T., Davila, J.M.,
et al. 2009, Sol. Phys., 259, 227

\bibitem[1965]{Harvey65} Harvey, G.A. 1965, JGR, 70, 2961

\bibitem[2004]{Hudson04} Hudson, H.S., \& Warmuth, A. 2004, ApJ, 614, L85

\bibitem[1997]{Kerdraon97} Kerdraon, A., \& Delouis, J.-M. 1997, in
Lect. Notes Phys. 483, Coronal Physics from Radio and Space Observations,
ed. G. Trottet, 192

\bibitem[2003]{Klassen03} Klassen, A., Pohjolainen, S., \& Klein, K.-L. 2003,
Sol. Phys., 218, 197

\bibitem[1999]{Klassen99} Klassen, A., Aurass, H., Klein, K.-L., Hofmann, A.,
\& Mann, G. 1999, A\&A, 343, 287

\bibitem[1999]{Klein99} Klein, K.-L., Khan, J.I., Vilmer, N., Delouis, J.-M.,
\& Aurass, H. 1999, A\&A, 346, L53

\bibitem[2009]{Liu09} Liu, Y., Luhmann, J.G., Bale, S.D., \& Lin, R.P. 2009,
ApJ, 691, L151

\bibitem[2008]{Magdalenic08} Magdaleni\'{c}, J., Vr\v{s}nak, B., Pohjolainen, S.,
et al. 2008, Sol. Phys. 253, 305

\bibitem[2010]{Magdalenic10} Magdaleni\'{c}, J., Marqu\'{e}, C., Zhukhov, A.N.,
Vr\v{s}nak, B., \v{Z}ic, T.  2010, ApJ, 718, 266

\bibitem[2000]{Maia00} Maia, D., Pick, M., Vourlidas, A., \& Howard, R. 2000,
ApJ, 528, L49

\bibitem[2002]{Mancuso02} Mancuso, S., Raymond, J.C., Kohl, J., et al.  2002, 
A\&A, 383, 267

\bibitem[1995]{Mann95} Mann, G., Classen, H.T., Aurass, H. 1995, A\&A, 295, 775

\bibitem[1996]{Mann96} Mann, G., Klassen, A., Classen, H.T., et al. 1996, A\&AS, 
119, 489

\bibitem[1962]{Maxwell62} Maxwell, A., \& Thompson, A.R 1962, ApJ, 135, 138

\bibitem[1985]{Nelson85} Nelson, G.J., \& Melrose, D.B. 1985, Solar 
Radiophysics, ed. D.J. McLean, \& N. Labrum (Cambridge University Press), 333 

\bibitem[1975]{Nelson75} Nelson, G.J., \& Robinson, R.D. 1975, Proc. Astr. Soc.
Australia, 2, 370

\bibitem[2008]{Nindos08} Nindos, A., Aurass, H., Klein, K.-L., \& Trottet, G.
2008, Sol. Phys., 253, 3

\bibitem[2009]{Ontiveros09} Ontiveros, V., \& Vourlidas, A. 2009, ApJ, 693, 267

\bibitem[2010]{Patsourakos10} Patsourakos, S., Vourlidas, A., \& Kliem, B.
2010, A\&A, 522, A100

\bibitem[2008]{Pick08} Pick, M., \& Vilmer, N. 2008, Astron. Astrophys. Rev.,
16, 1

\bibitem[2006]{Pohjolainen06} Pohjolainen, S., \& Lehtinen, N.J. 2006, A\&A,
449, 359

\bibitem[2007]{Pohjolainen07} Pohjolainen, S., van Driel-Gesztelyi, L.,
Culhane, J.L., Manoharan, P.K., \& Elliott, H.A. 2007, Sol. Phys., 244, 167

\bibitem[2008]{Pohjolainen08} Pohjolainen, S., Pomoell, J., \& Vainio, R. 2008, 
A\&A, 490, 357

\bibitem[2009]{Pothitakis09} Pothitakis, G., Preka-Papadema, P., Moussas, X.,
et al. 2009, in IAU Symp. 257, ed. N. Gopalswamy, D.F. Webb, 299

\bibitem[1982]{Priest82} Priest, E.R. 1982, Solar Magnetohydrodynamics
(Reidel, Dordrecht)

\bibitem[2000]{Raymond00} Raymond, J.C., Thompson, B.J., St. Cyr, O.C., et al. 
2000, Geophys. Res. Lett., 27, 1439

\bibitem[2007]{Reiner07} Reiner, M.J., Krucker, S., Gary, D.E., et al. 2007, 
ApJ, 657, 1107

\bibitem[1975]{Smerd75} Smerd, S.F., Sheridan, K.V., \& Stewart, R.T. 1975,
ApL, 16, 23

\bibitem[2005]{Shanmugaraju05} Shanmugaraju, A., Moon, Y.-J., Cho, K.-S.,
et al. 2005, Sol. Phys., 232, 87

\bibitem[2006]{Shanmugaraju06} Shanmugaraju, A., Moon, Y.-J., Cho, K.-S.,
Dryer, M., \& Umapathy, S. 2006, Sol. Phys., 233, 117

\bibitem[2000]{Sheeley00} Sheeley, N.R., Jr., Hakala, W.N., \& Wang, Y.-M.
2000, JGR, 105, 5081

\bibitem[2007]{Sheeley07} Sheeley, N.R., Jr., Warren, H.P., \& Wang, Y.-M.
2007, ApJ, 671, 926

\bibitem[1984]{Steinolfson84} Steinolfson, R.S. 1984, Sol. Phys., 94, 193

\bibitem[2006]{Subramanian06} Subramanian, K.R., \& Ebenezer, E. 2006,
A\&A, 451, 683

\bibitem[2008]{Temmer08} Temmer, M., Veronig, A.M., Vr\v{s}nak, B., et al.
2008, ApJ, 673, L95

\bibitem[1973]{Uchida73} Uchida, Y., Altschuler, M.D., \& Newkirk, G., Jr.
1973, Sol. Phys., 28, 495

\bibitem[2003]{Vourlidas03} Vourlidas, A., Wu, S.T., Wang, A.H., Subramanian,
P., \& Howard, R.A. 2003, ApJ, 598, 1392

\bibitem[2001a]{Vrsnak01a} Vr\v{s}nak, B. 2001a, JGR, 106, 25291

\bibitem[2001b]{Vrsnak01b} Vr\v{s}nak, B. 2001b, Sol. Phys. 202, 173 

\bibitem[2008]{Vrsnak08} Vr\v{s}nak, B., \& Cliver, E.W. 2008, Sol. Phys., 253, 215

\bibitem[1995]{Vrsnak95} Vr\v{s}nak, B., Ru\v{z}djak, V., Zlobec, P., \& Aurass,
H. 1995, Sol. Phys., 158, 331

\bibitem[2001]{VAM01} Vr\v{s}nak, B., Aurass, H., Magdaleni\'{c}, J., Gopalswamy, N.
2001, A\&A, 377, 321

\bibitem[2002]{Vrsnak02} Vr\v{s}nak, B., Magdaleni\'{c}, J., Aurass, H., Mann, G. 2002,
A\&A, 396, 673

\bibitem[2006]{Vrsnak06} Vr\v{s}nak, B., Warmuth, A., Temmer, M., et al. 2006, 
A\&A, 448, 739

\bibitem[2004]{Warmuth04} Warmuth, A., Vr\v{s}nak, B., Magdaleni\'{c}, J., Hanslmeier,
A., \& Otruba, W. 2004, A\&A, 418, 1117

\bibitem[2007]{White07} White, S.M., Mercier, C., Bastian, T., \& Bradley,
R. 2007, NRAO Newsletter, 110, 9

\bibitem[2005]{Williams05} Williams, D.R., T\"{o}r\"{o}k, T., D\'{e}moulin, P., 
van Driel-Gesztelyi, \& Kliem, B. 2005, ApJ, 628, L163

\bibitem[2001]{Zhang01} Zhang, J., Dere, K.P., Howard, R.A., Kundu, M.R., \&
White, S.M. 2001, ApJ, 559, 452

\bibitem[2004]{Zhang04} Zhang, J., Dere, K.P., Howard, R.A., \& Vourlidas, A.
2004, ApJ, 604, 420

\end{thebibliography}
\end{document}